\documentclass[letter]{aa}
\usepackage{graphicx,natbib}
\usepackage{inputenc}
\usepackage{txfonts}
\usepackage{color}

\begin{document} 

   \authorrunning{Charbonnel et al.}
   \titlerunning{Why the globular cluster NGC 6752 contains no sodium-rich second-generation AGB star}

   \title{Why the globular cluster NGC 6752 contains no sodium-rich second-generation AGB stars}

   \subtitle{}

   \author{Corinne Charbonnel\inst{1,2} William Chantereau\inst{1}, Thibaut Decressin\inst{1}, Georges Meynet\inst{1}, \and Daniel Schaerer\inst{1,2}
          }

   \institute{Geneva Observatory, University of Geneva, 51, ch. des Maillettes, CH-1290 Versoix, Switzerland\\
   \email{corinne.charbonnel@unige.ch} \and
   Institut de Recherche en Astrophysique et Plan\'etologie, CNRS UMR 5277, Universit\'e de Toulouse, 14, Av.E.Belin, 31400 Toulouse, France\\
             }

   \date{Received August 1, 2013; accepted August 28, 2013 }

  \abstract
   {Globular clusters host multiple stellar populations showing different sodium enrichments. These various populations can be observed along the main sequence, red giant and horizontal branch phases. Recently
   it was shown, however, that at least in the globular cluster NGC 6752, no sodium-rich stars are observed along the early asymptotic giant branch, posing an apparent problem for stellar evolution.}
  {We present  an explanation for this lack of sodium-rich stars in this region of the colour-magnitude diagram.}
  {We computed models for low-mass stars following the prediction of the so-called fast rotating massive stars model for the initial composition of second-generation stars. We studied the impact of different initial helium contents on the stellar lifetimes and the evolutionary path in the HertzsprungÐRussell diagram.}
  {We propose that the lack of sodium-rich stars along the early-AGB arises because sodium-rich stars were born with a high initial helium abundance, as predicted by the fast rotating massive stars scenario.  Helium-rich stars have much shorter lifetimes for a given initial mass than stars with a normal helium abundance, and above a cutoff initial helium abundance that slightly depends on the mass-loss rate on the RGB they do not go through the AGB phase and evolve directly into a white dwarf stage. 
  Within the fast rotating massive stars framework we obtained a cutoff in [Na/Fe] between the second-generation models evolving into the AGB phase and those skipping that phase between 0.18 and 0.4~dex, depending on the mass loss rate used during the red giant phase. In view of the uncertainties in abundance determinations, the cutoff obtained by the present model agrees well with the one inferred from
  recent observations of the cluster NGC 6752.}
  {The helium-sodium correlation needed to explain the lack of sodium-rich stars along the early-AGB of NGC 6752 corresponds to the one predicted by the fast rotating massive stars models. A crucial additional test of the model is the 
 distribution of stars with various helium abundances among main-sequence stars. Our model predicts that two magnitudes below the turnoff a very large percentage of stars, about 82\%, probably has a helium content lower than 0.275 in mass fraction, while only 5\% of stars are expected to have helium abundances greater than 0.4. 
  }
\maketitle

%

\section{Introduction}

 Long thought to be composed of coeval stars born out of homogeneous material, Galactic globular clusters are now known to host at least two generations of long-lived low-mass stars with different chemical properties \citep[e.g.][ and references therein]{2012A&ARv..20...50G}. First-generation stars have 
normal sodium abundance similar to that of halo stars. On the other hand, second-generation stars are identified thanks to their sodium overabundances. Both populations are found at the main-sequence turnoff, the subgiant branch, and the red giant branch (RGB) in all Milky Way globular clusters studied so far \citep[e.g.][]{2001A&A...369...87G,2009A&A...503..545L,2009A&A...505..139C}. At these evolution phases second-generation stars are counted to be about twice as numerous as their first-generation counterparts \citep{2010A&A...516A..55C}. 
 
However, a recent spectroscopic study \citep[][ hereafter C13]{2013Natur.498..198C} has revealed that the early asymptotic giant Bbranch (AGB) of the metal-poor globular cluster NGC 6752 hosts only stars with normal and slightly enhance  sodium abundance, while its RGB is populated by both sodium-normal and sodium-rich stars \citep{2007A&A...464..967C}. 
The result of C13 can be directly related to the deficiency of CN-strong AGB stars first discovered in NGC~6752 \citep{1981ApJ...244..205N}, which was subsequently reported for several other globular clusters \citep[see][ and references therein]{2010A&A...522A..77G,2012ASPC..458..205C}, although with strong cluster-to-cluster variations. It confirms the suggestion by \citet{2010A&A...522A..77G} that the most sodium-rich (and oxygen-poor) stars in globular clusters, which probably have a high helium and nitrogen content, could become AGB-manqu\'e because of their small mass on the horizontal branch (see \S 3).

Here we show that  the correlation between helium and sodium enhancement in the initial composition of second-generation low-mass stars that was predicted by the so-called {\it fast-rotating massive star polluter model} (hereafter FRMS) for NGC 6752 \citep{2007A&A...464.1029D} provides a very simple and natural explanation to the different sodium distributions observed on the RGB and AGB in this cluster. Contrary to C13, who interpreted the data as a failure for {\it {all}} second-generation stars to climb the AGB, 
we conclude that among the second generation, only stars born with initial helium and sodium abundances above a cutoff value evolve directly to the white dwarf phase after central helium-burning, 
while those with lower helium and sodium content climb the AGB in the same manner as their first generation counterparts. 
Note that because of the large uncertainties on sodium abundance determination (typical error bars of 0.2~dex), it is currently not possible to observationally distinguish first- and second-generation stars below the sodium cutoff value.

 
 \section{Helium-sodium correlation in the framework of the fast rotating massive stars scenario}

\subsection{General framework}

\begin{figure}
\begin{center}
\includegraphics[width=\columnwidth]{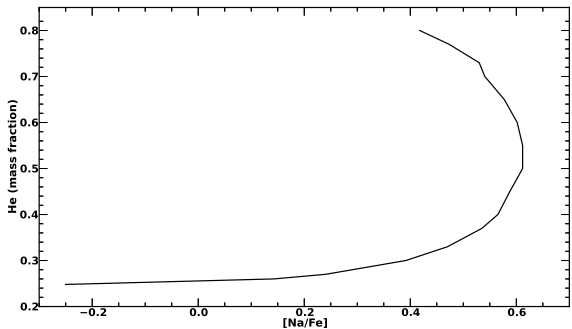}
\end{center}
\caption{Correlation predicted by the FRMS model for NGC~6752 for the initial helium and sodium abundances of second-generation stars \citep{2007A&A...464.1029D} that is adopted in our models of low-mass stars
}
\label{HeliumSodiumdistribution}
\end{figure}

It is now well established that second-generation globular cluster stars formed out of the sodium-rich ashes of hydrogen burning at high temperature that were ejected by more massive first-generation globular cluster stars and were diluted with interstellar gas \citep{2007A&A...470..179P}. Second-generation stars are therefore expected to start their life with a higher helium content than their first-generation counterparts. The impact of this helium enrichment on the extension of the horizontal branch has been extensively studied in the literature \citep[e.g.][]{2002A&A...395...69D,2010A&A...517A..81G,2010ApJ...708..698D}, and the connection with missing AGB stars in globular clusters has also been reported \citep{2010A&A...522A..77G}. 

However, the extent of the helium enrichment for a given sodium distribution strongly differs between the different polluter models. As we discuss here, this has serious implications because stars with different helium contents burn their nuclear fuel at different rates and are therefore expected to have different lifetimes. 

The nature of the polluters that have long since disappeared is debated \citep[see e.g.][]{2006A&A...458..135P}. Possible candidates are fast-rotating massive stars (FRMS) with masses larger than 25~M$_{\odot}$  \citep{2007A&A...464.1029D}, and massive (6-10~M$_{\odot}$) AGB stars \citep{2001ApJ...550L..65V}.  In both cases the sodium-rich ejecta are enriched in helium. 
In the AGB polluter model, helium-enrichement is due to the second dredge-up in intermediate-mass stars \citep{1997A&AS..123..241F,2007A&A...476..893S,2010MNRAS.407..854D,2013MNRAS.431.3642V}.  
It occurs before the thermal pulse phase on the AGB, when the competition between third dredge-up and hot-bottom-burning is supposed to enrich the stellar ejecta in sodium. 
Helium-enrichement of second-generation stars is thus limited and is
not directly correlated with the sodium content of the polluter yields. 
Consequently, the sodium spread observed in the cluster M4, which is quite similar to NGC 6752, is reproduced with limited helium enrichment almost identical for all second-generation stars 
\citep{2012MNRAS.423.1521D}. Therefore, and as shown by C13, 
RGB and AGB second-generation stars are expected to cover the same range in sodium abundances in this case, which is at odds with the observations in NGC 6752.  
To reconcile the AGB polluter model with the observations, C13 suggested that the mass loss rates during central helium-burning on the horizontal branch and latter phases is enhanced by an ad hoc factor of about 20.

In contrast, the FRMS pollution scenario predicts broad and correlated
spreads in both sodium and helium among second-generation low-mass stars
\citep{2007A&A...475..859D}. Consequently at a given age, stars with very
different helium and sodium abundances are expected to populate the
  different evolutionary phases. In particular, the observed sodium
  distribution is expected to be very different between the main sequence and RGB on
  one hand and the AGB on the other hand, as discussed below.

\subsection{Models for second-generation stars}

\subsubsection{Initial abundances}
We computed stellar evolution models with the code STAREVOL \citep{2000A&A...358..593S,2012A&A...543A.108L} for low-mass stars with various initial masses and helium abundances with a similar metallicity as NGC 6752 ([Fe/H]=-1.56; \citealt[][ C13]{1996AJ....112.1487H,2005A&A...440..901G}). 
For our computations we chose a uniform helium mass fraction of 0.248 for first-generation stars (close to the Big Bang abundance  
deduced from cosmological parameters obtained with Planck; \citealt{2013arXiv1303.5076P}), and a sodium abundance [Na/Fe]=-0.25 that corresponds to the observed value attributed to NGC~6752 first generation stars. For second-generation stars we adopted the correlation between initial helium and sodium abundances predicted by the FRMS polluter models for NGC 6752 RGB stars \citep{2007A&A...475..859D} that is shown in Fig.\ref{HeliumSodiumdistribution}. This distribution accounts for the dilution of H-burning ashes ejected by FRMS with pristine gas to reproduce the abundance variations of lithium in NGC 6752 \citep{2005A&A...441..549P}. The initial abundances of carbon, oxygen, nitrogen, magnesium, and aluminium are scaled accordingly. More details on the models will be presented in a forthcoming paper  (Chantereau et al., in prep.).

\subsubsection{Mass loss and other assumptions}
\begin{figure}
\begin{center}
\includegraphics[width=0.9\columnwidth]{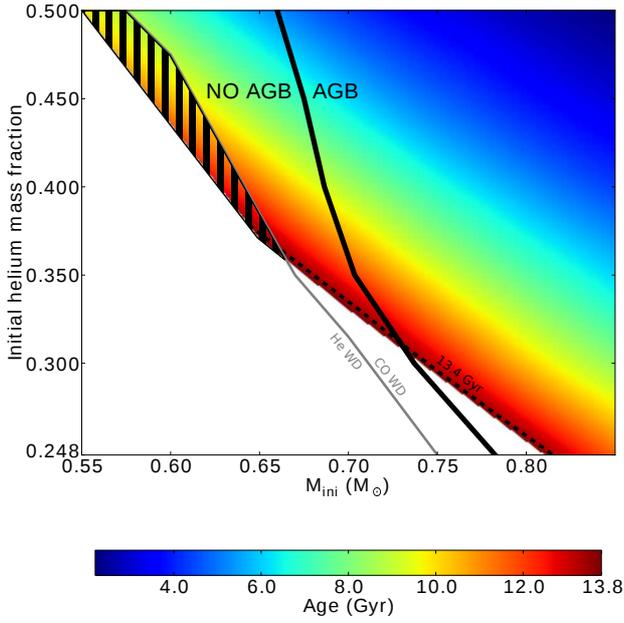}
\end{center}
\caption{
Lifetime and fate of stars versus initial mass and helium content at the
metallicity of NGC 6752. The colour code gives ages at the end of central
helium burning, or at the RGB tip for stars that fail to ignite helium
(dashed area) using the Reimers mass loss rate with $\eta=0.5$. The 13.4 Gyr isochrone is highlighted. The white area corresponds to lifetimes longer than the age of the Universe. The grey line separates stars that end up as helium or carbon-oxygen white dwarfs. The bold black line corresponds to the AGB ascension cutoff. It separates stars that climb the AGB (right-hand side) from the AGB-manqu\'e ones (left)
}
\label{Figure1}
\end{figure}

Mass loss during the RGB was accounted for following the Reimers prescription with $\eta$=0.5 \citep{1975MSRSL...8..369R}. 
Additional models were computed with $\eta$=0.65 for stars with an He abundance greater than 0.263 (see \S~\ref{ModelPredictions}). 
We neglected atomic diffusion and rotation-induced mixing; this simplification induces an uncertainty on the stellar lifetime of typically 0.2-0.5~Gyr, which is small compared to error bars on the current age estimate of NGC 6752 (13.4 $\pm$1.1 Gyr; \citealt{2003A&A...408..529G}). The FRMS pollution scenario predicts a maximum delay of 10 Myrs between the formation of the first and second stellar generations \citep{2013A&A...552A.121K}, which is negligible compared to the globular cluster age. Finally, the duration of the AGB phase is typically 0.1\% of the total stellar lifetime for the mass and helium domains we explore.  These age uncertainties can therefore be neglected in the discussion and we can safely consider that all post-helium-burning stars in NGC 6752 are along or very nearby the 13.4 Gyr isochrone (dashed black line in Fig.~\ref{Figure1}). 

\section{Model predictions}
\label{ModelPredictions}
The impact of initial mass and initial helium content on the lifetime and fate of cluster stars is illustrated in Fig.~\ref{Figure1}. 
The stars that are currently (i.e. at 13.4 Gyr) either at the main-sequence turnoff or are climbing the RGB have sodium abundances covering the range in [Na/Fe] 
observed at that phase in the cluster \citep[][ C13]{2007A&A...464..967C} and an initial helium mass fraction between 0.248 and 0.8 (see \S~\ref{discussion} for a discussion on the fraction of stars in various bins of helium abundance). 

\begin{figure}
\includegraphics[width=\columnwidth]{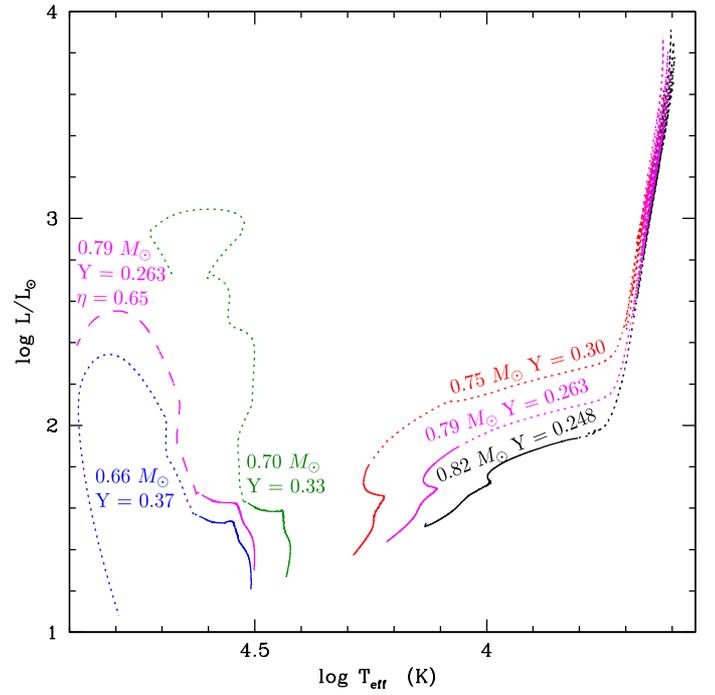}
\caption{Theoretical evolution path in the Hertzsprung-Russell diagram for stars in NGC 6752. Along individual tracks the solid line corresponds to the central helium-burning phase, and the dotted line to the more advanced phases (AGB or AGB-manqu\'e). All models are computed with a value of $\eta$=0.5 in the Reimers mass loss prescription, and we also show the prediction with $\eta$=0.65 for the 0.79~M$_{\odot}$, Y=0.263 model (long dashed). They correspond to a first-generation star (black) and to several second-generation stars (colours) that terminate helium burning at approximately 13.4 Gyr. The corresponding initial masses and helium abundances are given in the graph. The 0.82~M$_{\odot}$ star born with pristine helium as well as second-generation stars born with modest helium abundances evolve along the horizontal branch at a relatively low temperature and then all converge towards the AGB.  Instead second-generation stars born with higher helium abundances evolve at much higher effective temperatures on the horizontal branch and become AGB-manqu\'e}
\label{Figure3}
\end{figure}

We examine now the predictions for the abundance distribution on the AGB. 
Importantly, among the stars that burn helium at their centre, those with initial helium abundance above 0.31 (in mass fraction) and initial masses below 0.735 M$_{\odot}$ conclude their life without climbing the AGB, as can be seen in Fig.~\ref{Figure1}. 
These so-called AGB-manqu\'e (failed-AGB, \citealt{1990ApJ...364...35G}) lie along the post-helium-burning 13.4 Gyr isochrone in the domain between the grey and black lines in Fig.~\ref{Figure1}. 
They undergo central helium-burning at very high effective temperatures on the horizontal branch (Fig.~\ref{Figure3}) on the left side of the Grundahl jump. At that phase the stellar envelope surrounding the nuclear active core is extremely thin. As a consequence, once the helium fuel is exhausted at the centre, the energy released by the contraction of the core and by hydrogen shell-burning mainly diffuses across the thin, transparent helium-rich external layers and fails to swell them up.  Therefore the stars do not develop an extensive outer convection zone and do not move towards the AGB along the Hayashi track. Instead, they first evolve at a constant effective temperature towards higher luminosity before moving towards the white dwarf region (Fig.~\ref{Figure3}). 

We note that the AGB ascension cutoff values for the initial helium abundance and stellar mass are slightly sensitive to the mass loss rate adopted along the RGB. 
With the assumed FRMS helium-sodium distribution and using a value of 0.5 for the mass loss rate parameter $\eta$, the theoretical AGB ascension cutoff is obtained for a helium mass fraction of 0.31 which corresponds to an initial [Na/Fe] value of 0.4 dex. Thus the highest sodium abundance for second-generation stars that move towards the AGB is 0.4 dex in that case. This is higher than the upper envelope value of 0.18 dex observed for AGB stars in NGC 6752. 
On the other hand, using a value of 0.65 (instead of 0.5) is
sufficient to shift the theoretical AGB ascension cutoff along the 13.4Gyr
isochrone down to the observed [Na/Fe] cutoff value of 0.18. This can be
seen in Fig.~\ref{Figure3}, where we show the evolution paths for both
values of $\eta$ for the star that lies at the observational [Na/Fe] cutoff
of 0.18 dex on the 13.4 Gyr isochrone (i.e., initial mass and helium mass
fraction of 0.79~M$_{\odot}$ and 0.263). Depending on whether $\eta$ is 0.5 or 0.65, the model either moves towards the AGB or becomes an
AGB-manqu\'e.  Therefore we found that within the FRMS framework, a modest
increase in the mass loss along the RGB for the most helium-rich stars, compatible with modified Reimers-type
prescriptions \citep{2007A&A...465..593S} that propose a stronger
dependency of the mass-loss rate with stellar effective temperature than
the classical Reimers formula, is sufficient to achieve a perfect agreement
between the theoretical and observational sodium cutoff on the AGB. 
This is different from the attempt of C13 to fit the sodium data by assuming an increased mass loss rate by a factor of 20 on the horizontal branch for all second generation stars.

It is interesting to estimate the fraction of horizontal branch stars that evolve along the AGB depending on the value of $\eta$. 
Typically, for $\eta=0.5$ (a limiting value of He = 0.31), models show that $\sim$ 85\% climb the AGB, while a modest increase from $\eta = 0.5$ to 0.65 reduces this fraction to 25\%, which agrees reasonably well with the 30\% given by C13.

\section{Discussion and conclusions}
\label{discussion}
We showed that the correlation between the helium and sodium abundances expected from the FRMS models by \citet{2007A&A...464.1029D} for NGC~6752 provides a reasonable solution for the observed lack
of sodium-rich stars along the early-AGB in this globular cluster. The FRMS scenario predicts a high initial dispersion of helium abundances, and one may wonder whether this requirement agrees with the distributions of stars
in the observed colour magnitude diagram. For NGC 6752 a recent study by \citet[][ hereafter M13]{2013ApJ...767..120M} obtained that the observed colour dispersions at a given magnitude on the main sequence and on the RGB can be well fitted by a helium dispersion of only 0.03-0.04 in mass fraction, much lower than the helium dispersion predicted in our model. However, a higher helium dispersion than the one inferred by M13  for NGC 6752 can probably not be ruled out for the
following reasons: first, the chance of observing stars of a given initial helium abundance at a given magnitude and age decreases strongly for high helium values. Typically, 
starting from the initial distribution in helium given by \cite{2007A&A...464.1029D}, which reproduces the observed ratio 1:2 between the first and second generation of stars,
we have the following distribution of the number of stars as function of the helium abundance: 82.4\% with a helium
lower than 0.275, 12.6\% with a helium between 0.275 and 0.4, and 5\% with a helium between 0.4 and 0.7. These numbers are given for an age 
of 13.4 Gyr and are estimated at a 
position corresponding to two magnitudes below the turn off. 
Second, the dispersion obtained by M13 was deduced from shifts in colours of averaged positions of various main sequence and RGB lines, but there are stars that are bluer than the bluest averaged lines used by M13. 
This may indicate that there are stars with higher helium abundances than inferred by these authors. 
Third,  the link between colours and helium and/or other element abundances is by far not straightforward, since it requires computating model atmospheres and synthetic spectra that probably have their own uncertainties. 
Moreover, M13 explored only a relatively narrow set of parameters, and from their study one cannot exclude that higher helium abundances might be compatible with the observations. 
Thus, until comparisons become available for various atmosphere models and various codes that compute synthetic spectra for a large range of chemical compositions, we 
cannot discard the possibility that stars with higher helium abundances than those inferred by M13 are present.

According to C13, all stars bluer than the Grundahl jump in NGC 6752 (i.e. bluer than 4.1-4.3 in $\log T_{\rm eff}$) do not ascend the AGB. 
From Fig.~\ref{Figure3} we see that the limit in effective temperature between stars evolving into the AGB phase and those skipping this phase is between 4.1 (for $\eta$ equal to 0.65 instead of 0.5 for helium-rich stars) and 4.3
(for models with a constant value of $\eta$). Therefore our models are quite consistent with the limit in effective temperature suggested by C13. However, as noted above, this limit depends on the calibration process that allows
translating colours into effective temperatures and thus may present some uncertainties depending on which atmosphere and synthetic spectrum model is used.

Although the observed cutoff value [Na/Fe]$\sim$0.18 dex is somewhat uncertain (compare with the typical error bars of 0.2 dex usually quoted in the literature for sodium abundance determination), and in view on the uncertainties on the mass loss rate for evolved low-mass stars, we consider the agreement between the observations and the predictions of the current models very satisfactory. 

In conclusion, our results, obtained without fine-tuning the initial helium and sodium distribution or the physics of the stellar models, 
satisfactorily reconcile the different sodium patterns along the RGB and AGB of NGC 6752, and naturally explain the absence of AGB stars with high sodium abundances in this globular cluster. 
This indicates that the FRMS represents a reasonable solution for the origin of the chemical anomalies in globular clusters. As outlined above, some additional efforts are needed to
unambiguously constrain the actual dispersion in helium abundances within these systems. 

\begin{acknowledgements}
We acknowledge support from the Swiss National Science Foundation,  ESF-Eurogenesis, 
and the French Centre National de la Recherche Scientifique.  We are grateful to the referee A.Weiss for his constructive comments and to A.Maeder for careful reading of the manuscript.
\end{acknowledgements}

\end{document}